\documentclass{ws-p8-50x6-00}

\begin{document}

\title{Large $\mathrm{{\bf N_c}}$ physics from the lattice}

\author{M. Teper}

\address{Theoretical Physics, University of Oxford, 1 Keble Road,
Oxford OX1 3NP, UK\\
E-mail: teper@thphys.ox.ac.uk}

\maketitle

\abstracts{
I summarise what lattice methods can contribute to our understanding 
of the phenomenology of QCD at large $N_c$ and describe some recent 
work on the physics of SU($N_c$) gauge theories. These
non-perturbative calculations show that there is indeed 
a smooth $N_c\to\infty$ limit and that it is achieved by keeping
$g^2N_c$ fixed, confirming the usual diagrammatic analysis. 
The lattice calculations support the crucial assumption that the theory
remains linearly confining at large $N_c$. Moreover we see explicitly
that $N_c=3$ is `close to' $N_c=\infty$ for many physical 
quantities. We comment on the fate of topology and the deconfining
transition at  large $N_c$. We find that multiple confining
strings are strongly bound. The string tensions of these
$k$-strings are close to the  M(-theory)QCD-inspired conjecture 
that $\sigma_{k} \propto \sin(\pi k/N_c)$ as well as to
`Casimir scaling', $\sigma_{k} \propto k(N_c-k)$, with 
the most accurate recent calculations favouring the former.
We point out that closed $k$-strings provide a natural
way for non-perturbative effects to introduce $O(1/N_c)$
corrections into the pure gauge theory, in contradiction to
the conventional diagrammatic expectation. }

\section{Introduction}
\label{sec_introduction}

As we have seen at this meeting, large-$N_c$ arguments
\cite{largeN}
are very useful in illuminating many aspects of QCD.
Although we may not know very much about the detailed
physics at $N_c=\infty$, we can get a long way by assuming 
that there is a smooth large $N_c$ limit which is confining
and that $N_c=3$ is `close to' that limit. Analysis of the
colour flow in Feynman diagrams tells us that this limit is 
achieved by keeping constant the 't Hooft coupling, 
$\lambda \equiv g^2N_c$, and that the leading corrections are 
$O(1/N_c)$ in QCD and $O(1/N^2_c)$ in the gauge theory. 

The question I want to address in this talk is:
are the above assumptions correct and does the colour
flow counting survive if we go beyond diagrams to a
fully non-perturbative calculation? The technique I shall 
use is to discretise the theory onto a space-time lattice,
calculate various mass ratios via computer simulation, 
do this for a large enough
range of lattice spacings that one can confidently extrapolate
to the continuum limit; and finally, repeat the exercise for
a large enough range of $N_c$ that one can confidently control
the approach to $N_c=\infty$. 

I will do this for $SU(N_c)$ gauge theories with no quarks. 
These are calculations that can be -- and have been -- done on 
workstations. The states of this theory are purely
gluonic so one may call them `glueballs'. If we take the 
$N_c\to\infty$ limit of this glueball spectrum, what we obtain
is the glueball spectrum of $QCD_{N_c=\infty}$, since we expect
no mixing between glueballs and quarkonia at leading order in 
$1/N_c$ (at least for $m_q\not= 0$). The next step would be to
do $QCD_{N_c}$ in the `quenched' approximation, where all
quark vacuum bubbles are neglected. This can be regarded as a 
relativistic valence quark approximation to the theory. At any 
fixed $m_q\not= 0$ it has  $QCD_{N_c=\infty}$ as its 
$N_c\to\infty$ limit,
since quark vacuum bubbles do not appear at leading order
in $1/N_c$, and so it can be used to determine the quarkonium 
physics of that theory. Such a calculation would be interesting 
and should be possible using a Teraflop computer of the kind
that is becoming available to a number of lattice groups.
If however one wants to look at the chiral limit at
large $N_c$, then one needs to include quark loops and
these are calculations that are not for the near future.

In the next section I briefly remind you how one calculates
masses using lattice simulation and I give you an explicit
example to demonstrate that such calculations are indeed possible.
I then proceed to describe the results of a calculation 
\cite{blmtglue}
of the lightest few glueballs which shows that the approach
to $N_c=\infty$ is remarkably rapid: even SU(2) is `close to' 
SU($\infty$) for many quantities. These calculations provide 
some explicit evidence that linear confinement survives at 
large $N_c$ and that the limit is indeed achieved by keeping
$g^2N_c$ fixed. A much larger calculation of this kind
is now in progress
\cite{blmtuw}.
I then summarise the lattice results 
\cite{blmtstring,Pisa}
for the tensions of $k$-strings and what they imply for
various model/theoretic expectations. These string tension
calculations are also interesting because they provide an
explicit example of how non-perturbative physics may violate
the usual large $N_c$ diagrammatic colour-counting results.
Finally I briefly summarise what we are learning about the 
deconfining transition and about topology.

\section{Calculating Masses}
\label{sec_calc}

To calculate a mass we construct some operator $\phi(t)$
with the quantum numbers of the state and then use the
standard decomposition of the Euclidean correlator in
terms of energy eigenstates
\be
C(t) 
\equiv
\langle \phi^{\dagger}(t) \phi(0) \rangle
= 
\sum_n | \langle n | \phi | \Omega \rangle |^2
\exp \{- E_n t \}
\label{eqn_corrln}
\ee
where $|n\rangle$ are the energy eigenstates, with $E_n$ the 
corresponding energies, and $|\Omega\rangle$ is the vacuum state. 
To be able to evaluate the corresponding 
Euclidean Feynman Path Integrals
we discretise continuous space-time to a lattice and
truncate the infinite volume to a finite hypertorus. 
We now have a finite number of degrees of freedom
and can calculate Feynman Path Integrals using
standard Monte Carlo techniques. 

The lattice degrees of freedom are SU($N_c$) 
matrices that reside on the links of the lattice.
In our partition function the fields are weighted with
$\exp\{S\}$ where $S$ is the standard plaquette action
\be
S = - \beta \sum_p \biggl (1 - {1\over N_c} ReTr \, U_p \biggr ),
\label{eqn_action}
\ee
and $U_p$ is the ordered product of the matrices on
the boundary of the plaquette $p$. For smooth fields this
action reduces to the usual continuum action with
$\beta = 2N_c/g^2$. However the fields that dominate
the Feynman Path Integral are rough, all the way to the scale 
of the lattice spacing $a$. For these fields
we can define a running lattice coupling $g_L(a)$ which 
reduces in the continuum limit to 
a coupling $g(a)$ in our favourite scheme:
\be
\beta \equiv {{2N_c}\over{g_L^2(a)}}
\stackrel{a\to0}{\longrightarrow} {{2N_c}\over{g^2(a)}}
\label{eqn_beta}
\ee
So by varying the inverse lattice coupling $\beta$ we 
vary the lattice spacing $a$.

If we use a lattice action with reflection postivity, such as 
the simple plaquette action in eqn(\ref{eqn_action}),
then the decomposition in eqn(\ref{eqn_corrln}) remains valid,
except that now  $t=an_t$, so that
we obtain the energies from eqn(\ref{eqn_corrln}) as
$aE_n$ i.e. in units of the lattice spacing.

Having calculated some masses $am_i$ at a fixed value of $a$
we can remove lattice units
by taking ratios: $am_i/am_j = m_i/m_j$. This ratio
differs from the desired continuum value by lattice corrections.
For our action the functional form of the leading correction is
known to be $O(a^2)$. Thus for small enough $a$ we can 
extrapolate our calculated mass values
\be
{{m_i(a)} \over {m_j(a)}} =
{{m_i(0)} \over {m_j(0)}} 
+ c a^2 m^2_k(a)
\label{eqn_cont}
\ee
where $c$ depends on $i,j$ and $k$ and
the $a$-dependence of $m_k(a)$ will make differences at $O(a^4)$.
At this point we have
obtained the mass ratios of the continuum theory which is the
ultimate goal of our lattice calculations.

\begin	{figure}
\begin	{center}
\leavevmode
\setlength{\unitlength}{0.240900pt}
\ifx\plotpoint\undefined\newsavebox{\plotpoint}\fi
\sbox{\plotpoint}{\rule[-0.200pt]{0.400pt}{0.400pt}}%
\begin{picture}(1200,900)(0,0)
\font\gnuplot=cmr10 at 12pt
\gnuplot
\sbox{\plotpoint}{\rule[-0.200pt]{0.400pt}{0.400pt}}%
\put(425.0,194.0){\rule[-0.200pt]{4.818pt}{0.400pt}}
\put(400,194){\makebox(0,0)[r]{\ \ \ {$-1.5$}}}
\put(1105.0,194.0){\rule[-0.200pt]{4.818pt}{0.400pt}}
\put(425.0,413.0){\rule[-0.200pt]{4.818pt}{0.400pt}}
\put(400,413){\makebox(0,0)[r]{\ \ \ {$-1$}}}
\put(1105.0,413.0){\rule[-0.200pt]{4.818pt}{0.400pt}}
\put(425.0,631.0){\rule[-0.200pt]{4.818pt}{0.400pt}}
\put(400,631){\makebox(0,0)[r]{\ \ \ {$-0.5$}}}
\put(1105.0,631.0){\rule[-0.200pt]{4.818pt}{0.400pt}}
\put(425.0,850.0){\rule[-0.200pt]{4.818pt}{0.400pt}}
\put(400,850){\makebox(0,0)[r]{\ \ \ {$0$}}}
\put(1105.0,850.0){\rule[-0.200pt]{4.818pt}{0.400pt}}
\put(425.0,150.0){\rule[-0.200pt]{0.400pt}{4.818pt}}
\put(425,100){\makebox(0,0){$0$}}
\put(425.0,830.0){\rule[-0.200pt]{0.400pt}{4.818pt}}
\put(525.0,150.0){\rule[-0.200pt]{0.400pt}{4.818pt}}
\put(525,100){\makebox(0,0){$2$}}
\put(525.0,830.0){\rule[-0.200pt]{0.400pt}{4.818pt}}
\put(625.0,150.0){\rule[-0.200pt]{0.400pt}{4.818pt}}
\put(625,100){\makebox(0,0){$4$}}
\put(625.0,830.0){\rule[-0.200pt]{0.400pt}{4.818pt}}
\put(725.0,150.0){\rule[-0.200pt]{0.400pt}{4.818pt}}
\put(725,100){\makebox(0,0){$6$}}
\put(725.0,830.0){\rule[-0.200pt]{0.400pt}{4.818pt}}
\put(825.0,150.0){\rule[-0.200pt]{0.400pt}{4.818pt}}
\put(825,100){\makebox(0,0){$8$}}
\put(825.0,830.0){\rule[-0.200pt]{0.400pt}{4.818pt}}
\put(925.0,150.0){\rule[-0.200pt]{0.400pt}{4.818pt}}
\put(925,100){\makebox(0,0){$10$}}
\put(925.0,830.0){\rule[-0.200pt]{0.400pt}{4.818pt}}
\put(1025.0,150.0){\rule[-0.200pt]{0.400pt}{4.818pt}}
\put(1025,100){\makebox(0,0){$12$}}
\put(1025.0,830.0){\rule[-0.200pt]{0.400pt}{4.818pt}}
\put(1125.0,150.0){\rule[-0.200pt]{0.400pt}{4.818pt}}
\put(1125,100){\makebox(0,0){$14$}}
\put(1125.0,830.0){\rule[-0.200pt]{0.400pt}{4.818pt}}
\put(425.0,150.0){\rule[-0.200pt]{168.630pt}{0.400pt}}
\put(1125.0,150.0){\rule[-0.200pt]{0.400pt}{168.630pt}}
\put(425.0,850.0){\rule[-0.200pt]{168.630pt}{0.400pt}}
\put(200,500){\makebox(0,0){\Large{$\log C(t)$}}}
\put(775,25){\makebox(0,0){\Large{t}}}
\put(425.0,150.0){\rule[-0.200pt]{0.400pt}{168.630pt}}
\put(425,850){\usebox{\plotpoint}}
\put(415.0,850.0){\rule[-0.200pt]{4.818pt}{0.400pt}}
\put(415.0,850.0){\rule[-0.200pt]{4.818pt}{0.400pt}}
\put(475.0,788.0){\usebox{\plotpoint}}
\put(465.0,788.0){\rule[-0.200pt]{4.818pt}{0.400pt}}
\put(465.0,789.0){\rule[-0.200pt]{4.818pt}{0.400pt}}
\put(525.0,731.0){\rule[-0.200pt]{0.400pt}{0.482pt}}
\put(515.0,731.0){\rule[-0.200pt]{4.818pt}{0.400pt}}
\put(515.0,733.0){\rule[-0.200pt]{4.818pt}{0.400pt}}
\put(575.0,676.0){\rule[-0.200pt]{0.400pt}{0.723pt}}
\put(565.0,676.0){\rule[-0.200pt]{4.818pt}{0.400pt}}
\put(565.0,679.0){\rule[-0.200pt]{4.818pt}{0.400pt}}
\put(625.0,622.0){\rule[-0.200pt]{0.400pt}{0.964pt}}
\put(615.0,622.0){\rule[-0.200pt]{4.818pt}{0.400pt}}
\put(615.0,626.0){\rule[-0.200pt]{4.818pt}{0.400pt}}
\put(675.0,567.0){\rule[-0.200pt]{0.400pt}{1.686pt}}
\put(665.0,567.0){\rule[-0.200pt]{4.818pt}{0.400pt}}
\put(665.0,574.0){\rule[-0.200pt]{4.818pt}{0.400pt}}
\put(725.0,514.0){\rule[-0.200pt]{0.400pt}{1.927pt}}
\put(715.0,514.0){\rule[-0.200pt]{4.818pt}{0.400pt}}
\put(715.0,522.0){\rule[-0.200pt]{4.818pt}{0.400pt}}
\put(775.0,460.0){\rule[-0.200pt]{0.400pt}{2.891pt}}
\put(765.0,460.0){\rule[-0.200pt]{4.818pt}{0.400pt}}
\put(765.0,472.0){\rule[-0.200pt]{4.818pt}{0.400pt}}
\put(825.0,407.0){\rule[-0.200pt]{0.400pt}{4.095pt}}
\put(815.0,407.0){\rule[-0.200pt]{4.818pt}{0.400pt}}
\put(815.0,424.0){\rule[-0.200pt]{4.818pt}{0.400pt}}
\put(875.0,355.0){\rule[-0.200pt]{0.400pt}{5.541pt}}
\put(865.0,355.0){\rule[-0.200pt]{4.818pt}{0.400pt}}
\put(865.0,378.0){\rule[-0.200pt]{4.818pt}{0.400pt}}
\put(925.0,301.0){\rule[-0.200pt]{0.400pt}{7.468pt}}
\put(915.0,301.0){\rule[-0.200pt]{4.818pt}{0.400pt}}
\put(915.0,332.0){\rule[-0.200pt]{4.818pt}{0.400pt}}
\put(975.0,239.0){\rule[-0.200pt]{0.400pt}{10.600pt}}
\put(965.0,239.0){\rule[-0.200pt]{4.818pt}{0.400pt}}
\put(965.0,283.0){\rule[-0.200pt]{4.818pt}{0.400pt}}
\put(1025.0,177.0){\rule[-0.200pt]{0.400pt}{13.972pt}}
\put(1015.0,177.0){\rule[-0.200pt]{4.818pt}{0.400pt}}
\put(425,850){\circle*{12}}
\put(475,789){\circle*{12}}
\put(525,732){\circle*{12}}
\put(575,678){\circle*{12}}
\put(625,624){\circle*{12}}
\put(675,571){\circle*{12}}
\put(725,518){\circle*{12}}
\put(775,466){\circle*{12}}
\put(825,416){\circle*{12}}
\put(875,366){\circle*{12}}
\put(925,317){\circle*{12}}
\put(975,261){\circle*{12}}
\put(1025,206){\circle*{12}}
\put(1015.0,235.0){\rule[-0.200pt]{4.818pt}{0.400pt}}
\sbox{\plotpoint}{\rule[-0.500pt]{1.000pt}{1.000pt}}%
\put(425,836){\usebox{\plotpoint}}
\put(425.00,836.00){\usebox{\plotpoint}}
\put(439.16,820.84){\usebox{\plotpoint}}
\put(453.32,805.68){\usebox{\plotpoint}}
\put(467.48,790.52){\usebox{\plotpoint}}
\put(482.16,775.84){\usebox{\plotpoint}}
\put(496.32,760.68){\usebox{\plotpoint}}
\put(510.47,745.53){\usebox{\plotpoint}}
\put(524.63,730.37){\usebox{\plotpoint}}
\put(538.79,715.21){\usebox{\plotpoint}}
\put(552.95,700.05){\usebox{\plotpoint}}
\put(567.11,684.89){\usebox{\plotpoint}}
\put(581.79,670.21){\usebox{\plotpoint}}
\put(595.95,655.05){\usebox{\plotpoint}}
\put(610.11,639.89){\usebox{\plotpoint}}
\put(624.27,624.73){\usebox{\plotpoint}}
\put(638.43,609.57){\usebox{\plotpoint}}
\put(652.59,594.41){\usebox{\plotpoint}}
\put(666.75,579.25){\usebox{\plotpoint}}
\put(681.78,564.97){\usebox{\plotpoint}}
\put(695.93,549.80){\usebox{\plotpoint}}
\put(710.07,534.63){\usebox{\plotpoint}}
\put(724.22,519.46){\usebox{\plotpoint}}
\put(738.37,504.29){\usebox{\plotpoint}}
\put(752.52,489.12){\usebox{\plotpoint}}
\put(766.67,473.95){\usebox{\plotpoint}}
\put(781.34,459.33){\usebox{\plotpoint}}
\put(795.49,444.16){\usebox{\plotpoint}}
\put(809.63,428.99){\usebox{\plotpoint}}
\put(823.78,413.82){\usebox{\plotpoint}}
\put(837.93,398.65){\usebox{\plotpoint}}
\put(852.08,383.48){\usebox{\plotpoint}}
\put(866.23,368.31){\usebox{\plotpoint}}
\put(880.90,353.69){\usebox{\plotpoint}}
\put(895.05,338.52){\usebox{\plotpoint}}
\put(909.19,323.35){\usebox{\plotpoint}}
\put(923.34,308.18){\usebox{\plotpoint}}
\put(937.75,293.25){\usebox{\plotpoint}}
\put(951.91,278.09){\usebox{\plotpoint}}
\put(966.07,262.93){\usebox{\plotpoint}}
\put(980.75,248.25){\usebox{\plotpoint}}
\put(994.91,233.09){\usebox{\plotpoint}}
\put(1009.07,217.93){\usebox{\plotpoint}}
\put(1023.22,202.78){\usebox{\plotpoint}}
\put(1037.38,187.62){\usebox{\plotpoint}}
\put(1051.54,172.46){\usebox{\plotpoint}}
\put(1065.70,157.30){\usebox{\plotpoint}}
\put(1073,150){\usebox{\plotpoint}}
\end{picture}

\end	{center}
\caption{SU(4) correlation function for the $0^{++}$
glueball; exponential fit shown.}
\label{fig_corrln}
\end 	{figure}
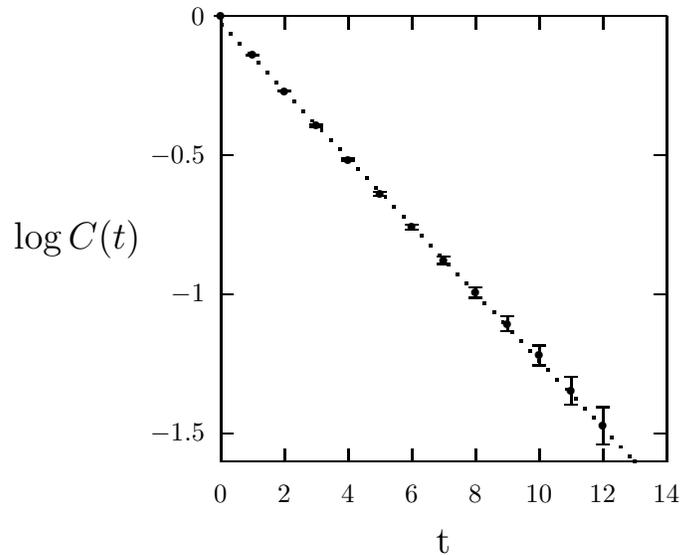

If we want to calculate the lightest mass using eqn(\ref{eqn_corrln})
then it is clear that we have to go to large enough $t$ that the
contribution of the excited states has died away and the
correlation function has acquired a simple exponential fall-off
with $t$. At large $t$, however,
the value of the correlation function becomes very small and
it is not obvious that a numerical approach, with finite errors,
will be accurate enough. To demonstrate that it can be, I show in
Fig.\ref{fig_corrln} the correlation function used to extract
the lightest SU(4) $J^{PC}=0^{++}$ glueball mass in an ongoing 
calculation
\cite{blmtuw}.
On this plot a simple exponential is a straight line and it
is clear that the corresponding mass can be determined very 
accurately. It is also clear that the simple exponential
decay already starts at small $t$. This means that the
operator we are using must be a good approximation to the
lightest glueball wavefunctional. This is no accident; it
has been obtained by a variational procedure which is an
important ingredient in the successful lattice calculation of
glueball masses, but one which I have no time to describe
further here
\cite{blmtglue}.

\vskip -0.05in

\begin	{figure}
\begin	{center}
\leavevmode
\setlength{\unitlength}{0.240900pt}
\ifx\plotpoint\undefined\newsavebox{\plotpoint}\fi
\sbox{\plotpoint}{\rule[-0.200pt]{0.400pt}{0.400pt}}%
\begin{picture}(1200,900)(0,0)
\font\gnuplot=cmr10 at 12pt
\gnuplot
\sbox{\plotpoint}{\rule[-0.200pt]{0.400pt}{0.400pt}}%
\put(350.0,150.0){\rule[-0.200pt]{4.818pt}{0.400pt}}
\put(325,150){\makebox(0,0)[r]{\ \ \ {$0$}}}
\put(1105.0,150.0){\rule[-0.200pt]{4.818pt}{0.400pt}}
\put(350.0,306.0){\rule[-0.200pt]{4.818pt}{0.400pt}}
\put(325,306){\makebox(0,0)[r]{\ \ \ {$1$}}}
\put(1105.0,306.0){\rule[-0.200pt]{4.818pt}{0.400pt}}
\put(350.0,461.0){\rule[-0.200pt]{4.818pt}{0.400pt}}
\put(325,461){\makebox(0,0)[r]{\ \ \ {$2$}}}
\put(1105.0,461.0){\rule[-0.200pt]{4.818pt}{0.400pt}}
\put(350.0,617.0){\rule[-0.200pt]{4.818pt}{0.400pt}}
\put(325,617){\makebox(0,0)[r]{\ \ \ {$3$}}}
\put(1105.0,617.0){\rule[-0.200pt]{4.818pt}{0.400pt}}
\put(350.0,772.0){\rule[-0.200pt]{4.818pt}{0.400pt}}
\put(325,772){\makebox(0,0)[r]{\ \ \ {$4$}}}
\put(1105.0,772.0){\rule[-0.200pt]{4.818pt}{0.400pt}}
\put(350.0,150.0){\rule[-0.200pt]{0.400pt}{4.818pt}}
\put(350,100){\makebox(0,0){$0$}}
\put(350.0,830.0){\rule[-0.200pt]{0.400pt}{4.818pt}}
\put(544.0,150.0){\rule[-0.200pt]{0.400pt}{4.818pt}}
\put(544,100){\makebox(0,0){$0.05$}}
\put(544.0,830.0){\rule[-0.200pt]{0.400pt}{4.818pt}}
\put(738.0,150.0){\rule[-0.200pt]{0.400pt}{4.818pt}}
\put(738,100){\makebox(0,0){$0.1$}}
\put(738.0,830.0){\rule[-0.200pt]{0.400pt}{4.818pt}}
\put(931.0,150.0){\rule[-0.200pt]{0.400pt}{4.818pt}}
\put(931,100){\makebox(0,0){$0.15$}}
\put(931.0,830.0){\rule[-0.200pt]{0.400pt}{4.818pt}}
\put(1125.0,150.0){\rule[-0.200pt]{0.400pt}{4.818pt}}
\put(1125,100){\makebox(0,0){$0.2$}}
\put(1125.0,830.0){\rule[-0.200pt]{0.400pt}{4.818pt}}
\put(350.0,150.0){\rule[-0.200pt]{186.697pt}{0.400pt}}
\put(1125.0,150.0){\rule[-0.200pt]{0.400pt}{168.630pt}}
\put(350.0,850.0){\rule[-0.200pt]{186.697pt}{0.400pt}}
\put(100,500){\makebox(0,0){\Large{${{m[0^{++}]}\over{\surd\sigma}}$}}}
\put(737,25){\makebox(0,0){\Large{$a^2\sigma$}}}
\put(350.0,150.0){\rule[-0.200pt]{0.400pt}{168.630pt}}
\put(898.0,546.0){\rule[-0.200pt]{0.400pt}{3.854pt}}
\put(888.0,546.0){\rule[-0.200pt]{4.818pt}{0.400pt}}
\put(888.0,562.0){\rule[-0.200pt]{4.818pt}{0.400pt}}
\put(724.0,595.0){\rule[-0.200pt]{0.400pt}{4.095pt}}
\put(714.0,595.0){\rule[-0.200pt]{4.818pt}{0.400pt}}
\put(714.0,612.0){\rule[-0.200pt]{4.818pt}{0.400pt}}
\put(604.0,620.0){\rule[-0.200pt]{0.400pt}{4.818pt}}
\put(594.0,620.0){\rule[-0.200pt]{4.818pt}{0.400pt}}
\put(594.0,640.0){\rule[-0.200pt]{4.818pt}{0.400pt}}
\put(502.0,656.0){\rule[-0.200pt]{0.400pt}{5.300pt}}
\put(492.0,656.0){\rule[-0.200pt]{4.818pt}{0.400pt}}
\put(898,554){\circle*{12}}
\put(724,603){\circle*{12}}
\put(604,630){\circle*{12}}
\put(502,667){\circle*{12}}
\put(492.0,678.0){\rule[-0.200pt]{4.818pt}{0.400pt}}
\sbox{\plotpoint}{\rule[-0.500pt]{1.000pt}{1.000pt}}%
\put(350,713){\usebox{\plotpoint}}
\put(350.00,713.00){\usebox{\plotpoint}}
\put(369.92,707.32){\usebox{\plotpoint}}
\put(389.85,701.68){\usebox{\plotpoint}}
\put(409.73,695.82){\usebox{\plotpoint}}
\put(429.48,689.63){\usebox{\plotpoint}}
\put(449.32,683.67){\usebox{\plotpoint}}
\put(469.07,677.48){\usebox{\plotpoint}}
\put(488.99,671.75){\usebox{\plotpoint}}
\put(508.75,665.50){\usebox{\plotpoint}}
\put(528.60,659.52){\usebox{\plotpoint}}
\put(548.59,654.03){\usebox{\plotpoint}}
\put(568.47,648.13){\usebox{\plotpoint}}
\put(588.32,642.17){\usebox{\plotpoint}}
\put(608.10,635.97){\usebox{\plotpoint}}
\put(627.95,630.01){\usebox{\plotpoint}}
\put(647.69,623.83){\usebox{\plotpoint}}
\put(667.66,618.25){\usebox{\plotpoint}}
\put(687.39,611.89){\usebox{\plotpoint}}
\put(707.27,606.02){\usebox{\plotpoint}}
\put(727.26,600.53){\usebox{\plotpoint}}
\put(747.09,594.48){\usebox{\plotpoint}}
\put(766.94,588.52){\usebox{\plotpoint}}
\put(786.74,582.36){\usebox{\plotpoint}}
\put(806.57,576.36){\usebox{\plotpoint}}
\put(826.47,570.57){\usebox{\plotpoint}}
\put(846.37,564.74){\usebox{\plotpoint}}
\put(866.08,558.34){\usebox{\plotpoint}}
\put(885.98,552.51){\usebox{\plotpoint}}
\put(905.97,547.01){\usebox{\plotpoint}}
\put(925.75,540.81){\usebox{\plotpoint}}
\put(945.60,534.85){\usebox{\plotpoint}}
\put(965.41,528.74){\usebox{\plotpoint}}
\put(985.23,522.69){\usebox{\plotpoint}}
\put(1005.18,517.06){\usebox{\plotpoint}}
\put(1025.08,511.22){\usebox{\plotpoint}}
\put(1044.79,504.83){\usebox{\plotpoint}}
\put(1064.68,498.99){\usebox{\plotpoint}}
\put(1084.63,493.34){\usebox{\plotpoint}}
\put(1104.45,487.30){\usebox{\plotpoint}}
\put(1124.26,481.19){\usebox{\plotpoint}}
\put(1125,481){\usebox{\plotpoint}}
\end{picture}

\end	{center}
\caption{The lightest SU(4) scalar glueball mass, $m_{0^{++}}$, 
expressed in units of the string tension, $\sigma$, plotted 
against the latter in lattice units. The $a\to 0$ continuum 
extrapolation, using a leading lattice correction, is shown.}
\label{fig_g0cont}
\end 	{figure}
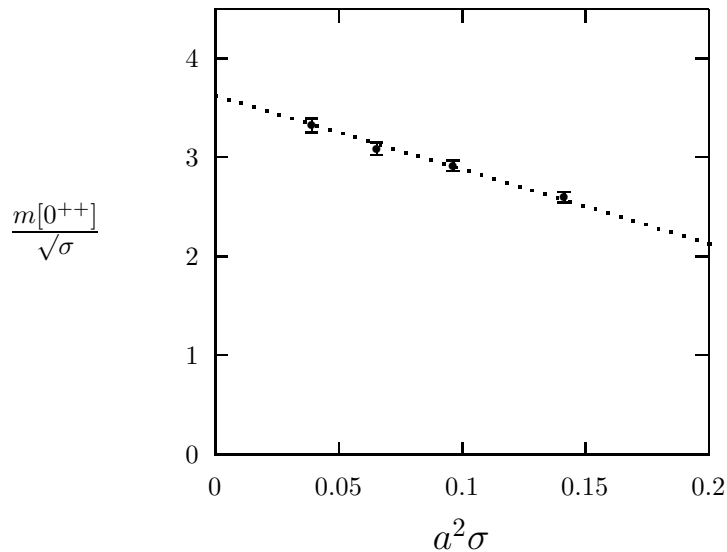

We have just seen that lattice calculations of masses are 
indeed possible. But are they accurate enough to permit a
controlled extrapolation to the continuum limit? To
demonstrate that the answer to this question is yes, I plot in
Fig.\ref{fig_g0cont} the (preliminary) $0^{++}$ masses obtained 
from the same calculation
\cite{blmtuw}
at four different values of $a$. The masses have been
expressed in terms of the confining string tension, $a^2\sigma$,
which has also been also calculated, and the ratio is plotted against
$a^2\sigma$. As we note from eqn(\ref{eqn_cont}) the leading
lattice correction is $O(a^2)$ which means that the
continuum extrapolation at sufficiently small $a$ will
be a straight line -- as shown in the plot. This provides
an example of a typical continuum extrapolation.

\vskip -0.05in

\section{SU($\mathrm{\bf{N_c}}$) Glueball Masses}
\label{sec_sunglue}

\begin	{figure}
\begin	{center}
\epsfig{figure=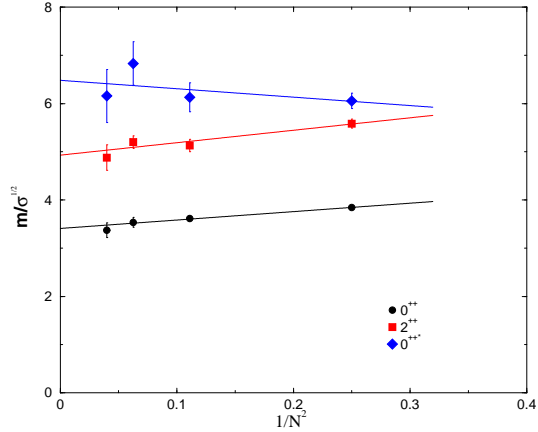, angle=270, width=7cm} 
\end	{center}
\caption{The mass of the lightest scalar glueball, $m_{0^{++}}$, 
.... expressed in units of the string tension, $\sigma$, is plotted 
against $1/N^2_c$. The $N_c\to \infty$ extrapolation is shown.}
\label{fig_sunmass}
\end 	{figure}

In 
\cite{blmtglue}
we calculated the lightest and first excited $0^{++}$ glueball
masses and the lightest tensor $2^{++}$ glueball mass. We took
the ratio to the string tension and extrapolated to the
continuum limit as described in Section~\ref{sec_calc}. We did 
this for SU(2), SU(3), SU(4) and SU(5) gauge theories. In 
Fig.\ref{fig_sunmass} I plot these continuum mass ratios
against $1/N^2_c$. We expect the leading correction
at large $N_c$ to be $O(1/N^2_c)$,
\be
\left.  {{m_i} \over {m_j}} \right |_{N_c} =
\left.  {{m_i} \over {m_j}} \right |_{\infty} 
+ {c_{ij} \over N_c^2},
\label{eqn_largeN}
\ee
which is a simple straight line on our plot. Remarkably,
as we see in  Fig.\ref{fig_sunmass}, the mass ratios
for all values of $N_c$ can be described by just the
leading correction and the corresponding coefficients
are modest in magnitude.

This shows us that there is indeed a smooth
large $N_c$ limit with a finite confining string tension.
Moreover for these quantities SU(3) is clearly close to 
SU($\infty$). Indeed, so is SU(2). That is to say,
SU($N_c$) gauge theories are close to SU($\infty$)
for all values of $N_c$.

\section{'t Hooft Coupling}
\label{sec_coupling}

We have seen that there is a smooth large-$N_c$ limit.
Is it achieved by keeping  constant the 't Hooft coupling, 
$\lambda \equiv g^2 N_c$, as suggested by the
standard analysis of diagrams 
\cite{largeN}?
In D=2+1 the coupling $g^2$ has dimensions of mass and
the question is simply whether  $g^2N_c/\surd\sigma$
goes to a non-zero finite constant as $N_c\to\infty$.
The answer is found to be yes
\cite{mtd3}.
Here in D=3+1 the coupling runs and is dimensionless.
The question therefore becomes
\cite{blmtglue,mtoldN}:
is the smooth large $N_c$
limit achieved by keeping fixed  the running
't Hooft coupling, as defined on some scale $l$ that 
is fixed in units of some quantity that partakes of 
the smooth large-$N_c$ limit, such as the string tension? 
To test this we use eqn(\ref{eqn_beta}) which tells is 
that a suitable defintion of a running 't Hooft coupling is 
\be
\lambda_I(a) =  g_I^2(a) N_c 
= {{2N_c^2}\over{\beta \langle ReTr \, U_p /N_c \rangle}}
\label{eqn_lambdaI}
\ee
The extra factor involving the plaquette is a standard mean-field 
(or tadpole) improved version of $\beta$ and the naive $\lambda(a)$ 
we would derive from it. Such improvements are customary
because the naive lattice coupling is known to be very poor 
in the sense of having very large higher order corrections.

\begin	{figure}
\begin	{center}
\epsfig{figure=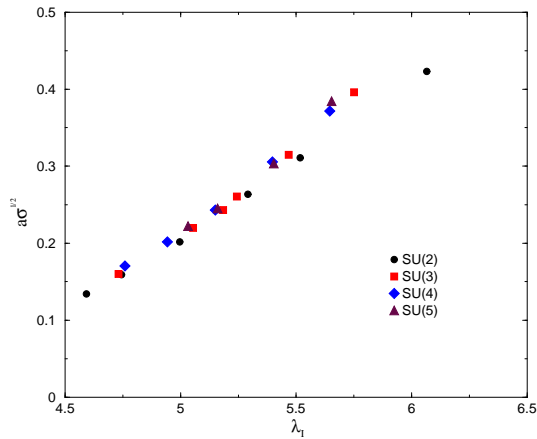, angle=270, width=7cm} 
\end	{center}
\caption{'t Hooft coupling at the scale $a$ in physical units.}
\label{fig_coupling}
\end 	{figure}

We extract $\lambda_I(a)$ and $a\surd\sigma$ for various values
of $a$. The latter expresses $a$ in physical units so that
a plot of $\lambda_I(a)$ against $a\surd\sigma$ is a plot
of how the coupling runs. If the large $N_c$ limit
requires a fixed 't Hooft coupling then we would expect
that such plots tend to a fixed curve as  $N_c\to\infty$.
As we see in Fig~\ref{fig_coupling} not only does this seem 
to be the case, but the limit is already achieved  at the
smallest non-trivial values of $N_c$.

\section{k-Strings}
\label{sec_kstrings}

We can consider confining strings between static colour charges
in arbitrary representations. However gluon screening means 
that the effective representation can be changed 
dynamically. It is therefore useful to label charges
by their transformation properties under the centre of the
group since gluons transform trivially under the centre.
Suppose the source  transforms by a factor of $z^k$ under a 
global gauge transformation $z$ belonging to the centre $Z_{N_c}$ 
of the SU($N_c$) group. Call the lightest confining string
joining sources in this class the $k$-string. The usual string
between quarks is the $k=1$ string. What is the tension 
$\sigma_k$ of such a string as a function of $k$ and $N_c$?
There are some conjectures (see
\cite{blmtstring} 
for details). For example, a form
\be
{{\sigma_{\mathit{k}}}\over{\sigma_1}}
=
{{\sin \frac{k\pi}{N_c}}\over{\sin \frac{\pi}{N_c}}} .
\label{eqn_mqcd}
\ee
has been conjectured in an M(-theory)QCD approach to QCD
\cite{MQCD}. 
Another relevant example is the old Casimir scaling 
\cite{cs}
hypothesis
\be
{{\sigma_{\mathit{k}}}\over{\sigma}}
= 
{{k(N_c-k)}\over{N_c-1}}
\label{eqn_casimir}
\ee
as well. (Note we use $\sigma\equiv\sigma_1$ from now on.)

\begin	{figure}
\begin	{center}
\epsfig{figure=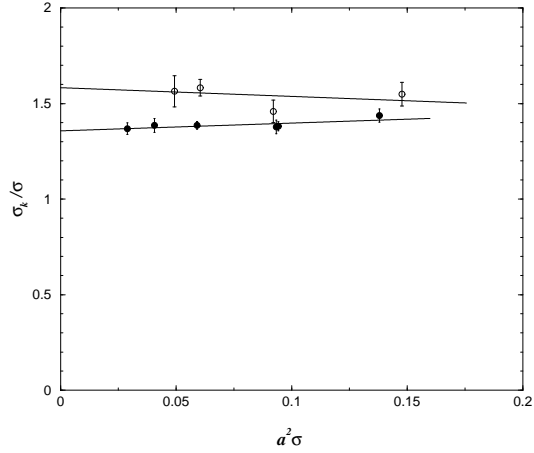, angle=270, width=7cm} 
\end	{center}
\caption{$k=2$ string tensions in SU(4) ($\bullet$) 
and SU(5) ($\circ$).}
\label{fig_kstring}
\end 	{figure}

In Fig~\ref{fig_kstring} I show the lattice values of 
$\sigma_{k=2}/\sigma$ as a function of $a^2\sigma$
obtained in our
\cite{blmtstring} 
recent SU(4) and SU(5) calculations.
(With continuum extrapolations.) One sees that 
$\sigma_{k=2}/\sigma \ll 2$ i.e. non-trivial strongly
bound $k$-strings do indeed exist. Moreover the 
continuum value lies between the Casimir scaling and
MQCD conjectures, which are numerically very similar.

A more recent and more accurate calculation 
\cite{Pisa} 
in SU(4) and SU(6) favours the MQCD conjecture. On the other 
hand our preliminary anisotropic lattice calculations
\cite{blmtuw} 
seem to favour an intermediate value.

\section{Corrections to the $\mathrm{\bf{N_c=\infty}}$ Limit?}
\label{sec_limitN}

Suppose that the $k$-string tension is eventually found to
satisfy Casimir scaling, as in eqn(\ref{eqn_casimir}).
Consider a $k$-string wrapped around a spatial hypertorus of
length $l$. This represents an energy eigenstate of the
finite volume Hamiltonian with mass $m_k = l \sigma_k$
with corrections of $O(1/l^2)$ that we can neglect for
large enough $l$. Assuming eqn(\ref{eqn_casimir}) we
obtain
\be
m_{\mathit{k}} = l \sigma_k 
=
l\sigma  {{k(N_c-k)}\over{N_c-1}}
=
l\sigma \{ k - {{k(k-1)}\over{N_c}} + \ldots \}  .
\label{eqn_stringNc}
\ee
Consider the $N_c$-dependence of this, keeping
$l$ fixed in units of, say, $\sigma$. We see that
the leading correction is $O(1/N_c)$. This contradicts
the usual diagrammatic result that the leading correction
in SU($N_c$) gauge theories should be $O(1/N^2_c)$. 

Perhaps we should not regard a string of fixed $k$ as being 
the `same' state as $N_c$ is varied. Then the above would
not worry us. But suppose that (part of) the glueball
spectrum arises from closed strings of flux, as for example
in the Isgur-Paton flux tube model
\cite{ip}. 
As pointed out in
\cite{mtrj} 
one can form such loops out of $k$-strings as well as out
of fundamental strings, leading to sectors of states
that are scaled in mass by $\sigma_k/\sigma$.
Eqn(\ref{eqn_stringNc}) tells us that
such glueball states will then have  $O(1/N_c)$
rather than  $O(1/N^2_c)$ corrections as
$N_c \to \infty$.

Of course we do not yet know whether $k$-strings
satisfy Casimir scaling or not. (Note that
the same issue arises in D=2+1.) But the general point
is that we have an explicit example of how non-perturbative
effects -- string formation -- might lead to a violation
of the usual colour counting rules. This is interesting
whether or not reality chooses to make use of this possibility.

\section{Conclusions}
\label{sec_conc}

I have not had time to discuss topology. Here one finds 
\cite{blmtglue}
that the SU($\infty$) topological susceptibility is
non-zero and not very different from the SU(3) one.
This is important for our understanding 
\cite{eta}
of how the $\eta^\prime$ gets its large mass. Moreover
fluctuations that are unambiguously instanton-like 
disappear from the vacuum as $N_c$ grows.
I have also not discussed the deconfining transition:
the nature of this transition is being actively
investigated
\cite{highT,blmtglue}.

What I have shown in this talk, using fully non-perturbative 
calculations, is that the large-$N_c$ limit is smooth, 
confining and is achieved precociously for many physical 
quantities. Not only is $N_c=3$ close to $N_c=\infty$ 
but so is  $N_c=2$. There are new stable strings at
larger $N_c$ and their string tensions are intrigueingly close 
to both the MQCD and Casimir scaling conjectures.

One obtains the $N_c\to\infty$ limit by keeping fixed 
the 't Hooft coupling. This is as expected. Not
expected was the observation that the $k$-strings
provide, in principle, an explicit avenue by which states 
can acquire anomalous $O(1/N_c)$ corrections.

Lattice calculations which will make our knowledge of 
SU($N_c$) gauge theories much more extensive are under way.

\section*{Acknowledgments}

I have learned a great deal from the talks and
various discussions in this `interdisciplinary' 
meeting; for example the observations made in
Section \ref{sec_limitN} came to me while discussing 
my lattice calculations with Simon Dalley, Matt Strassler 
and Aneesh Manohar. My thanks to the organizers for
inviting me.

\end{document}